\documentclass[journal = jctcce]{achemso}
\setkeys{acs}{articletitle = true}

\usepackage{graphicx}
\usepackage{float}
\usepackage{amsmath}
\usepackage{amssymb}
\usepackage{babel}
\usepackage{subcaption}
\usepackage{mathrsfs,framed,esint,slashed,braket,bm}
\usepackage{color}
\usepackage{braket}
\usepackage{placeins}
\usepackage{verbatim}
\usepackage{multirow}
\usepackage[toc,page]{appendix}
\usepackage[normalem]{ulem}

\newcommand{\wf}{wavefunction}

\newcommand{\be}{\begin{equation}}
\newcommand{\ee}{\end{equation}}

\title{The rovibrational Aharonov--Bohm effect}

\author{Jonathan I. Rawlinson}
\email{J.Rawlinson@leeds.ac.uk}
\affiliation{School of Mathematics, University of Leeds, Leeds, LS2 9JT, UK
}
\altaffiliation{MTA-ELTE Complex Chemical Systems Research Group, P.O. Box 32, H-1518 Budapest 112, Hungary}

\author{Csaba F\'abri}
\affiliation{Laboratory of Molecular Structure and Dynamics, 
Institute of Chemistry, ELTE E\"otv\"os Lor\'and University, 
P\'azm\'any P\'eter s\'et\'any 1/A, H-1117 Budapest, Hungary}
\altaffiliation{MTA-ELTE Complex Chemical Systems Research Group, P.O. Box 32, H-1518 Budapest 112, Hungary}

\author{Attila G. Cs\'asz\'ar}
\affiliation{Laboratory of Molecular Structure and Dynamics, 
Institute of Chemistry, ELTE E\"otv\"os Lor\'and University, 
P\'azm\'any P\'eter s\'et\'any 1/A, H-1117 Budapest, Hungary}
\altaffiliation{MTA-ELTE Complex Chemical Systems Research Group, P.O. Box 32, H-1518 Budapest 112, Hungary}

\begin{document}



\linespread{1.0}
\begin{abstract}
Another manifestation of the Aharonov--Bohm effect is introduced to chemistry,
in fact to nuclear dynamics and high-resolution molecular spectroscopy.
As demonstrated, the overall rotation of a symmetric-top molecule influences
the dynamics of an internal vibrational motion in a way that is analogous to the 
presence of a solenoid carrying magnetic flux. 
To a good approximation, the low-energy rovibrational energy-level structure
of the quasistructural molecular ion H$_5^+$ can be understood entirely in terms of 
this effect.
\end{abstract}

\newpage
\section{Introduction}
More than 60 years ago, following the footsteps of Ehrenberg and Siday,\cite{49EhSi}
Aharonov and Bohm \cite{59AhBo,61AhBo} provided a 
detailed analysis of the significance of electromagnetic potentials in quantum theory.
They made the then surprising claim that in the quantum domain these potentials 
have a physical and measurable effect on charged particles even in regions where 
the magnetic field vanishes.
The findings of Aharonov and Bohm were verified experimentally. \cite{tonomura1986evidence,tonomura1982observation} 
While at the beginning the statements of these studies about fields and potentials
may have seemed just counterintuitive curiosities of quantum theory,
over the years the Ehrenberg--Siday--Aharonov--Bohm 
(usually simply referred to as the Aharonov--Bohm) effect, and its various analogues,
found application in a number of different fields of molecular sciences. 
Of particular importance to this study is the so-called 
molecular Aharonov--Bohm (MAB) effect \cite{mead1979determination,mead1980molecular,mead1980electronic} 
within Born--Oppenheimer theory,\cite{27BoOp,54BoHu}
in which the conical intersection seam,\cite{63HeLo,99Yarkony} corresponding to 
degenerate electronic states, has an effect on the nuclear motion analogous 
to the presence of a solenoid carrying magnetic flux.  
The implications of the MAB effect for both bound states \cite{mead1980molecular,mead1980electronic}
and scattering events \cite{17Zygelman} have been explored.

We are aware of no indication of the significance of the Aharonov--Bohm effect
in nuclear dynamics unrelated to the MAB effect.
In this paper we introduce the `rovibrational Aharonov--Bohm effect',
occuring in symmetric-top molecules with a separable vibrational degree of freedom (dof).
As shown below, the `rotational' dynamics influence the 
`vibrational' dynamics in a way analogous to the coupling of the selected vibrational
motion to the field of a magnetic solenoid. 
We investigate in detail whether this analogy between rovibrational nuclear dynamics 
and electromagnetic phenomena can help us to understand high-resolution molecular spectra
and the related dynamics of molecules. 

The treatment of overall rotations coupled to large-amplitude internal motions
(sometimes referred to as contortions \cite{bunker2006molecular}) 
has been the subject of intensive studies in high-resolution molecular
spectroscopy.\cite{32Nielsen,59LiSw,70HoBuJo,72BuSt,72Pickett,74HoBu,77BuLa,94HoKlGo,03SzCsSaOr,bunker2006molecular,11Bauder}
Perhaps the most famous among the models is the pioneering 
Hougen--Bunker--Johns (HBJ) model,\cite{70HoBuJo}
describing the rotational-contortional dynamics of triatomic molecules 
with a large-amplitude dof, the bending.
There exist reviews \cite{59LiSw,94HoKlGo,bunker2006molecular} on the topic of
rotation-contortion-vibration Hamiltonians
(the contortions include internal rotation, pseudorotation, inversion, torsion, and bending)
and their utilization in the understanding of high-resolution molecular spectra.
The different Hamiltonians derived, for example,
the approximate principal-axis, internal-axis,
and rho-axis ones, are relevant to the present discussion, as they contain
different forms of kinetic energy and rotational-vibrational (Coriolis) coupling
terms obtained \emph{via} different rotational transformations.
Nevertheless, to the best of our knowledge,
the central idea of the present paper, the `electromagnetic analogy' 
of such Hamiltonians leading to the rovibrational Aharonov--Bohm effect was not
considered in these studies.

As an example of a molecular system exhibiting the rovibrational Aharonov--Bohm effect,
the low-energy nuclear dynamics of the H$_{5}^{+}$ molecular cation is considered.
First-principles characterization of the dynamics of H$_{5}^{+}$,
as well as of its deuterated derivatives, has been achieved and it
offered a number of considerable surprises.\cite{14FaSaCs,15SaFaSzCs,16SaCs}
As shown below, the rovibrational Aharonov--Bohm effect dominates 
the low-energy nuclear dynamics of this quasistructual \cite{20CsFaSa} molecular ion, 
which is approximately a symmetric top with a nearly-free torsional motion.

\section{The electromagnetic analogy} 

Let us start our discussion with the standard spectroscopic four-dimensional (4D)
rotation-contortion model Hamiltonian \cite{32Nielsen,59LiSw,84GoCo,11Bauder,20CsFaSa}
\begin{equation}
    \hat{H} = \hat{H}_{\rm v} + \hat{T}_{\rm r}^{\rm RR} + \hat{T}_{\rm rv} 
    = (a \hat{p}_{\phi}^2 + \hat{V}) + [B\hat{J}^2 + (A-B) \hat{J}^2_z] - \alpha \hat{p}_\phi \hat{J}_z,
\label{eq:genmodham}
\end{equation}
describing the dynamics of a symmetric-top molecule with a torsional ($\phi$, period $2\pi$)
degree of freedom (dof) coupled to the rotational dofs.
In Eq. \eqref{eq:genmodham}, $\hat{p}_{\phi} = -\textrm{i} \frac{\partial}{\partial \phi}$,
$a$ depends on the moments of inertia of the internal rotor and of the molecule 
about the molecular symmetry axis, the potential energy $\hat{V}$ depends
solely on $\phi$ and typically has multiple minima,
the three components of the overall body-fixed angular momentum $\hat{J}$
are denoted by $\hat{J}_i$, $A$ and $B$ are the effective rotational constants
of the symmetric top, and $\alpha$ is the rovibrational coupling strength.
The various operators appearing in $\hat{H}$ of Eq.~(\ref{eq:genmodham})
have been grouped, as usual, into the operators $\hat{H}_{\rm v}$ 
(vibrational Hamiltonian, containing only $\phi$ derivatives), 
$\hat{T}_{\rm r}^{\rm RR}$ (rotational Hamiltonian, containing only rotational derivatives),
and a rotation-vibration coupling term $\hat{T}_{\rm rv}$.

However, this is not the only possible way to group the terms of Eq.~(\ref{eq:genmodham}).
For example, some authors 
incorporate the rotation-vibration coupling 
into the vibrational term to give the rearranged Hamiltonian  \cite{32Nielsen,03SzCsSaOr,11Bauder}
\begin{equation}
    \hat{H}  = a \left(\hat{p}_{\phi}-\frac{\alpha\hat{J}_z}{2a}\right)^2 + \hat{V} + [B\hat{J}^2 + (A-\frac{\alpha^2}{4a}-B) \hat{J}^2_z].
    \label{toy}
\end{equation}

A common next step is to perform the so-called Nielsen transformation,\cite{32Nielsen}
which leads to the apparent elimination of the rovibrational coupling.
We will discuss the Nielsen transformation shortly. 
For now, we note that $K$, related to the projection of the angular momentum
onto the molecule-fixed $z$ axis, is a good quantum number. 
Setting $\hat{J}_z=K$ in Eq.~(\ref{toy}), 
we get that 
\begin{equation}
    \hat{H}  = \hat{H}^K_\mathrm{tor} + E_{\mathrm{rot}},
    \label{tor-rot}
\end{equation}
and so the Hamiltonian is given by the sum of a `torsional' Hamiltonian  
\begin{equation}
    \hat{H}^K_\mathrm{tor} = a \left(\hat{p}_{\phi}-\frac{\alpha K}{2a}\right)^2 + \hat{V} 
    \label{abtor}
\end{equation}
and a rotational kinetic energy contribution, given by
\begin{equation}
    E_\mathrm{rot}=BJ\left(J+1\right) + (A-\frac{\alpha^2}{4a}-B) K^2.
\end{equation}

We must point out that $\hat{H}^K_\mathrm{tor}$, 
which we are calling the `torsional' Hamiltonian,
in fact depends on the rotational quantum number $K$ through the appearance of the operator $\hat{p}_{\phi}-\frac{\alpha K}{2a}$. 
Essentially, $\hat{H}^K_\mathrm{tor}$ takes the usual kinetic plus potential form,
except for the fact that the torsional momentum $\hat{p}_\phi$ has been `modified' 
by the replacement $\hat{p}_\phi\to\hat{p}_{\phi}-\frac{\alpha K}{2a}$. 

The key insight is that this is precisely the modification one would make 
in order to couple the torsional motion to a magnetic field,
where $\frac{\alpha }{2a}$ plays the role of an electromagnetic vector potential
and $K$ is formally identified with the electric charge. 
In fact, the Hamiltonian in Eq.~(\ref{abtor}) is of exactly the same form 
as the Hamiltonian for a particle confined to a ring encircling a solenoid 
which is carrying magnetic flux,\cite{merzbacher1962single} just as in the famous Aharonov--Bohm effect.
It is as if the rotational dofs induce an effective magnetic field 
which couples to the torsional motion. 
We call this the \emph{rovibrational Aharonov--Bohm effect}.
It can be viewed as a special case, for symmetric tops, of the more general influence 
of rovibrational coupling on vibrational motion through so-called 
non-Abelian gauge fields (for more information on this general picture, see Refs.  \citenum{Shapere1987,Shapere1989,GUICHARDET1984,Littlejohn1997,Rawlinson2020a,rawlinson2019coriolis}).

As a result, the effect of rovibrational coupling on the torsional motion can be understood 
by analogy with the familiar effects of a magnetic field. 
To the best of our knowledge, this is a new perspective on the well-studied Hamiltonian of 
Eq.~(\ref{eq:genmodham}). 
As we will see, this electromagnetic analogy allows us to get deeper insight into the 
rovibrational level structure, and reveals properties which are obscured by the usual separation 
of nuclear-motion Hamiltonians into rotational, vibrational, and rovibrational terms.

\subsection{Changes of embedding}
Next, let us
consider the effect of a change of embedding of the molecule-fixed axes. 
Suppose that, for our original embedding, we have some rovibrational eigenstate $\chi_K\left(\phi\right)|J K M\rangle$ with the torsional \wf\  $\chi_K$ satisfying
\begin{equation}
    \left(\hat{H}^K_\mathrm{tor}+E_\mathrm{rot} \right)\chi_K=E\chi_K.
    \label{old}
\end{equation}

Imagine now that we change our embedding choice.
We only consider embeddings where the body-fixed $z$-axis is aligned with the symmetry axis of the molecule.
At each $\phi$, the old and new embedding must be related by a rotation 
about the body-fixed $z$-axis by some angle $\theta\left(\phi\right)$. 
With respect to the new embedding, 
the torsional \wf\ $\chi_K$ will become $\tilde{\chi}_K$, where
\begin{equation}
    \chi_K=\exp\left(-i\theta\left(\phi\right)K\right)\tilde{\chi}_K.
    \label{newtor}
\end{equation}
Substituting this relation into Eq.~(\ref{old}) gives
\begin{equation}
    a \left(\hat{p}_{\phi}-\left(\frac{\alpha}{2a}+\theta'\left(\phi\right)\right)K\right)^2\tilde{\chi}_K + \hat{V} \tilde{\chi}_K + E_{\rm rot}\tilde{\chi}_K=E\tilde{\chi}_K,
\end{equation}
and, by comparison with Eq.~(\ref{tor-rot}), we see that the rotational-energy contribution
looks exactly the same in the new embedding, and the only change to the torsional Hamiltonian is the shift
\begin{equation}
    \frac{\alpha }{2a}\rightarrow\frac{\alpha }{2a}+\theta'\left(\phi\right)
    \label{shift}
\end{equation}
of the electromagnetic vector potential by the $\phi$-derivative of $\theta$.
We recognise this as the usual transformation law for an electromagnetic vector potential
under a gauge transformation. 
Thus, we see that, within the electromagnetic analogy, changes of embedding 
correspond to electromagnetic gauge transformations.

These considerations reveal that splitting the Hamiltonian into an 
electromagnetically-coupled vibrational part and a rotational part,
as in Eq.~(\ref{toy}), has a significant advantage:
\emph{the rotational part becomes independent of the embedding choice}.
This is to be contrasted with the traditional splitting into vibrational 
plus rotational plus rovibrational terms [shown in Eq.~(\ref{eq:genmodham})],
in which the coefficients in the rotational term (the so-called effective rotational constants)
depend on the choice of embedding. 
The price one pays for this embedding-independence of the rotational term is that 
the vibrational term depends on the embedding. 
However, this dependence is only through the inclusion of the electromagnetic vector potential,
which simply transforms by a gauge transformation under changes of embedding.
In particular, it is obvious from our way of splitting up the Hamiltonian 
that the rovibrational energy levels are independent of the choice of embedding 
(as they should be). 
This is because of the well-known fact that the energy levels of a system 
coupled to an electromagnetic field are invariant under gauge transformations.
In the traditional way of writing the rovibrational Hamiltonian, 
it is far from obvious that the energy levels do not depend on the choice of embedding.
In other words, the traditional way of writing the Hamiltonian obscures 
the embedding-independence (or the \emph{gauge invariance}) of the rovibrational energy levels.

\subsection{Eliminating rotation-vibration coupling}
The electromagnetic analogy gives us insight into the important issue 
of the ``elimination'' of rotation-vibration coupling.\cite{21SaPoSzCs} 
Recall that, under a change of embedding specified by $\theta\left(\phi\right)$,
the vector potential transforms as in Eq.~(\ref{shift}). If we can choose our embedding so that the vector potential becomes zero, then clearly the torsional motion will be separated from the rotational motion. This seems possible, since we can just choose $\theta\left(\phi\right)=-\frac{\alpha\phi}{2a}$ and then Eq.~(\ref{shift}) tells us that the vector potential will become 
\begin{equation}
    \frac{\alpha }{2a}\rightarrow\frac{\alpha }{2a}+\left(-\frac{\alpha\phi}{2a}\right)'=\frac{\alpha }{2a}-\frac{\alpha}{2a}=0; 
\end{equation}
thus, with respect to this new embedding choice, the torsional Hamiltonian becomes 
\begin{equation}
    \hat{H}^K_\mathrm{tor} \rightarrow a \hat{p}_{\phi}^2 + \hat{V},
\end{equation}
which no longer depends on the rotational quantum number $K$.

Strictly speaking, however, $\theta\left(\phi\right)=-\alpha\phi/2a$ 
does not give a valid embedding transformation. 
To see this, first recall that the torsional coordinate $\phi$ is $2\pi$-periodic. 
On the other hand, the rotation angles $\theta\left(0\right)$ and $\theta\left(2\pi\right)$
relating the old and new embeddings at $\phi=0$ and $\phi=2\pi$ are not the same,
but differ by an angle $\theta\left(2\pi\right)-\theta\left(0\right)=-\alpha\pi/a$.
In other words, $\theta$ does not give a single-valued choice of embedding. 
As a consequence, the torsional \wf\ in the new embedding [see Eq.~(\ref{newtor})],
$\tilde{\chi}_K$, satisfies
\begin{equation}
    \tilde{\chi}_K\left(2\pi\right)=\exp\left(-i\alpha\pi K/a\right)\tilde{\chi}_K\left(0\right),
    \label{bcs}
\end{equation}
and so we see that, while the original torsional \wf\  $\chi_K$ was $2\pi$-periodic in $\phi$,
it becomes necessary to allow for modified boundary conditions, Eq.~(\ref{bcs}),
for the new torsional \wf\  $\tilde{\chi}_K$. 
These new boundary conditions `compensate' for the multi-valuedness of the new embedding.

For example, for a molecule composed of two identical rotors 
(for which it turns out that $\alpha/a=1$---see the following section for an example)
we have that
\begin{equation}
 \tilde{\chi}_K\left(2\pi\right)=\exp\left(-iK\pi \right)\tilde{\chi}_K\left(0\right)=\left(-1\right)^{K}\tilde{\chi}_K\left(0\right), 
 \label{antiper}
\end{equation}
or in other words that $\chi_K$ satisfies periodic/anti-periodic boundary conditions
for $K$ even/odd. 
So we are able to eliminate the vector potential, \emph{if we are prepared} 
to allow non-periodic boundary conditions on the torsional \wf.

This `multi-valued' change of embedding which eliminates the vector potential,
together with the modified boundary conditions which are required to accomodate it,
is essentially the well-known Nielsen transformation referred to earlier.
It is also analogous to elimination of the Mead--Truhlar--Berry (MTB) vector potential \cite{mead1979determination} in the context of the molecular Aharonov--Bohm effect, 
in which the resulting multi-valued nature of the electronic \wf\  leads to 
modified boundary conditions for the nuclear \wf. 
Specifically, the nuclear \wf\ picks up a minus sign upon encircling 
a conical intersection of potential energy surfaces, 
much like how the torsional \wf\ in Eq.~(\ref{antiper}) picks up 
a minus sign when $\phi$ goes from $0$ to $2\pi$. 
This analogy with conical intersections will be elaborated on in section~3.3.

The necessity to modify the boundary conditions,  as in Eq.~(\ref{bcs}),
is arguably an unpleasant feature of the Nielsen transformation. 
It is also unnatural from the perspective of the electromagnetic analogy, 
in which gauge transformations ought to be single-valued.
We might ask, therefore, whether it is possible to eliminate the vector potential
\emph{without} having to introduce modified boundary conditions for $\chi_K$. 
In other words, can we construct some single-valued change of embedding 
which transforms the vector potential away? 
The electromagnetic analogy makes it clear that the answer is generally `no'. 
This is because the electromagnetic vector potential $\alpha/2a$ corresponds 
to the presence of a  magnetic flux and, as is well known in the context of 
the traditional Aharonov--Bohm effect, this magnetic flux generally has 
a non-trivial effect on the quantum energy levels. 
It cannot simply be transformed away. 
The only exception is when $\alpha/2a=n\in\mathbb{Z}$ is an integer, since then the single-valued embedding transformation $\theta\left(\phi\right)=-n\phi$ yields
\begin{equation}
    \frac{\alpha }{2a}\rightarrow\frac{\alpha }{2a}+\left(-n\phi\right)'=n-n=0,  
\end{equation}
showing that the effect of the corresponding magnetic flux is 
no different to if there were no magnetic field at all.
This is a \emph{flux quantization} condition, as is familiar in the Aharonov--Bohm effect
[see, \emph{e.g.}, Eq.~(16) in Ref.~\citenum{merzbacher1962single}].

To summarize, the vector potential (and, in turn, the coupling between rotations and vibrations)
cannot generally be transformed away by a change of embedding. 
If one is prepared to work with multi-valued embedding transformations, 
then in a sense the vector potential can be eliminated, 
but with the price that the torsional \wf\ is no longer $2\pi$-periodic in $\phi$. 
This corresponds to the Nielsen transformation. 
In what follows, we opt to work with single-valued gauge transformations 
as this corresponds to the usual perspective taken in electromagnetic theory,
and preserves the interpretation of rovibrational coupling in terms of 
the presence of a magnetic flux.

\section{Application to H$_5^+$}

In this section we show how the rovibrational Aharonov--Bohm effect arises
in the context of the unusual rotation-torsion dynamics of the molecular cation H$_5^+$.

\begin{figure}[t!]
\noindent \begin{centering}
\includegraphics[scale=0.475]{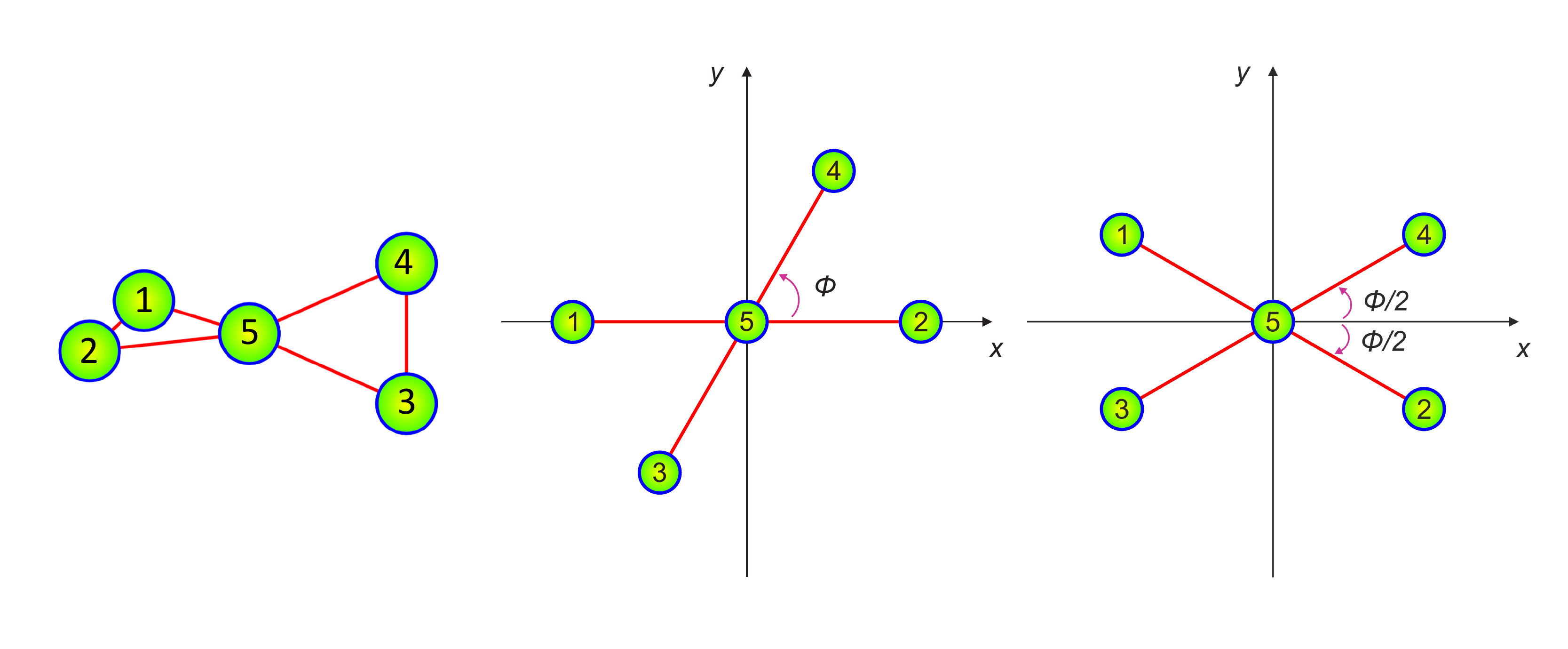}
\par\end{centering}
\caption{Equilibrium structure of the H$_5^+$ molecular ion (left panel) and definitions 
of the geometric (GE, middle panel) and bisector (BE, right panel) embeddings,
whereby $\phi$ corresponds to the torsional coordinate.}
\label{fig:Emb}
\end{figure}

\subsection{1D($\phi$) torsion model}
Our starting point is the reduced-dimensionality 1D($\phi$) torsion 
model \cite{14FaSaCs,20CsFaSa,15SaFaSzCs} developed to understand the low-energy
quantum dynamics of H$_{5}^{+}$. The equilibrium structure of H$_{5}^{+}$ is depicted in the left panel of Fig.~\ref{fig:Emb}. We choose to work in the so-called
geometric embedding (GE) of molecule-fixed axes,\cite{15SaFaSzCs}
see Fig.~\ref{fig:Emb},
with respect to which the positions of the five hydrogen atoms are
[see Eq.~(1) of Ref.~\citenum{15SaFaSzCs}]
\begin{align}
\mathbf{r}_{1}&=1/2\left(-1,0,R\right)^{T} \nonumber \\
\mathbf{r}_{2}&=1/2\left(1,0,R\right)^{T} \nonumber \\
\mathbf{r}_{3}&=1/2\left(-\cos\phi,-\sin\phi,-R\right)^{T} \\
\mathbf{r}_{4}&=1/2\left(\cos\phi,\sin\phi,-R\right)^{T} \nonumber \\
\mathbf{r}_{5}&=\left(0,0,0\right)^{T}, \nonumber
\end{align}
which depend on a torsional coordinate $\phi$, which has period $2\pi$, and $R$ is the
distance between the midpoints of the two H$_2$ units. 
Note that here, and throughout this section, we work in units where $\hbar=m_{\rm H}=r=1$,
where $m_{\rm H}$ is the mass of the hydrogen atom and $r$ is the distance
between the two hydrogens forming a rotor.

The time-independent Schrödinger equation for the 1D($\phi$) torsion model takes the form
\begin{equation}
2\left(\hat{p}_{\phi}-\frac{1}{2}\hat{J}_{z}\right)^{2}\psi^{\rm }+\sum_{i,j}^{}\frac{1}{2}\hat{J}_{i}\left(M^{-1}\right)_{ij}\hat{J}_{j}\psi^{\rm }+V\left(\phi\right)\psi^{\rm }+V_{2}\left(\phi\right)\psi^{\rm }=E\psi^{\rm }.
\label{rot-tor}
\end{equation}
In previous studies this equation has been split into separate rotational, vibrational and rovibrational contributions but here we have grouped the terms as suggested in the mathematical physics literature.\cite{Littlejohn1997} In Eq.~(\ref{rot-tor}), $\hat{p}_{\phi}=-\textrm{i}\partial/\partial\phi$ 
is the momentum conjugate to the torsional coordinate $\phi$ 
and the $\hat{J}_{i}$ are the components of the body-fixed angular momentum,
while $M^{\rm}$ is the moment of inertia tensor with respect
to the GE molecule-fixed frame, governing the motion along the three 
rotational dofs, and has the explicit form
\begin{equation}
M=\begin{pmatrix}R^{2}+\frac{1}{2}\sin^{2}\phi & -\frac{1}{2}\cos\phi\sin\phi & 0\\
-\frac{1}{2}\cos\phi\sin\phi & \frac{1}{2}+R^{2}+\frac{1}{2}\cos^{2}\phi & 0\\
0 & 0 & 1
\end{pmatrix}.
\end{equation}
Note that we have allowed for a torsional potential $V\left(\phi\right)$ 
in Eq.~(\ref{rot-tor}), which on physical grounds is assumed to satisfy
\begin{equation}
V\left(\phi\right)=V\left(-\phi\right)=V\left(\phi+\pi\right)
\end{equation}
for all $\phi$ (recall that $\phi$ has period $2\pi$). 
In addition, the extrapotential term $V_{2}\left(\phi\right)$ 
is a quantum contribution coming from the $\phi$-dependence of the rovibrational 
$\mathbf{G}$ matrix,\cite{bunker2006molecular} which in this case takes the explicit form
\begin{equation}
V_{2}\left(\phi\right)=\frac{8\left(1+8R^{2}+8R^{4}\right)\cos2\phi-\left(7+\cos4\phi\right)}{16\left(\left(1+2R^{2}\right)^{2}-\cos^{2}\phi\right)^{2}}.
\end{equation}

At this point let us pause to interpret the various terms in Eq.~(\ref{rot-tor}). 
Broadly speaking, the first term on the left-hand side should be thought
of as the torsional kinetic energy; the second term is the rotational
kinetic energy; the third term is the torsional potential energy,
which has been modified by the extrapotential term $V_{2}\left(\phi\right)$.
Note that our grouping of terms in Eq.~(\ref{rot-tor}) is different to the usual conventions
adopted in the nuclear-dynamics literature.
For example, note the appearance of the operator $\hat{p}_\phi-\frac{1}{2}\hat{J}_z$ 
in the first term, reminiscent of the electromagnetic coupling considered 
in the previous section. 
We will now make the connection to the previous section, and to the Aharonov--Bohm effect, more explicit.

\subsection{Symmetric-top approximation}

For applications to H$_{5}^{+}$, it is a very good approximation
to take $R$ large.\cite{14FaSaCs}
So let us expand in $1/R$.
To leading order,
\begin{equation}
V_{2}\left(\phi\right)=\frac{\cos2\phi}{4R^{4}}
\end{equation}
and 
\begin{equation}
M^{-1}=\begin{pmatrix}\frac{1}{R^{2}} & \frac{\sin2\phi}{4R^{4}} & 0\\
\frac{\sin2\phi}{4R^{4}} & \frac{1}{R^{2}} & 0\\
0 & 0 & 1
\end{pmatrix}.
\end{equation}
Discarding terms which are of order $\frac{1}{R^{4}}$ or higher,
$M^{-1}$ becomes diagonal and $V_{2}$ vanishes and so Eq.~(\ref{rot-tor}) reduces to
\begin{equation}
2\left(\hat{p}_{\phi}-\frac{1}{2}\hat{J}_{z}\right)^{2}\psi+\left[\frac{1}{2R^{2}}\hat{J}^{2}+\left(\frac{1}{2}-\frac{1}{2R^{2}}\right)\hat{J}_{z}^{2}\right]\psi+V\left(\phi\right)\psi=E\psi.
\label{limit}
\end{equation}

We call this (large-$R$) approximation the \emph{symmetric-top approximation},
since the rotational kinetic energy term is now that of a symmetric top. 
In fact, the Hamiltonian in Eq.~(\ref{limit}) is of exactly the same form 
as the Hamiltonian of Eq.~(\ref{toy}) considered in the previous section, 
with the particular values 
\begin{equation}
a=2, ~ \alpha=2, ~ B=\frac{1}{2 R^2}, ~ A=1.
\label{vals}
\end{equation}
So, by the results of the previous section, the rovibrational energy levels are of the form
\begin{equation}
    E=E^K_\mathrm{tor}+E_{\mathrm{rot}},
\end{equation}
where, in this case, the rotational energy contribution is
\begin{equation}
     E_\mathrm{rot}=\frac{1}{2 R^2}J\left(J+1\right) + \left(\frac{1}{2}-\frac{1}{2 R^2}\right) K^2
\end{equation}
and the torsional energy contribution is an eigenvalue of the torsional Hamiltonian
\begin{equation}
    \hat{H}^K_\mathrm{tor} = 2 \left(\hat{p}_{\phi}-\frac{K}{2}\right)^2 + \hat{V}.
    \label{tor}
\end{equation}
Note that the electromagnetic coupling $\hat{p}_{\phi}-\frac{K}{2}$ 
appearing in the torsional Hamiltonian of Eq.~(\ref{tor}) takes a particularly simple form,
due to the special value of the ratio $\frac{\alpha}{a}=1$ in this case.

We can now take advantage of the electromagnetic analogy to better understand
the torsional Hamiltonian of Eq.~(\ref{tor}), 
which is simply the Hamiltonian for a charged particle, confined to a ring,
encircling a solenoid carrying magnetic flux. 
The first insight that this analogy gives us is that \emph{the torsional energy levels only depend on whether $K$ is even or odd}, and not on the precise value of the integer $K$.
We prove this now.

Suppose that $E_{\mathrm{tor}}$ is an energy eigenvalue of $\hat{H}_{\mathrm{tor}}^{K}$,
with $\psi$ a corresponding eigenstate:
\begin{equation}
\hat{H}_{\mathrm{tor}}^{K}\psi=E_{\mathrm{tor}}\psi.
\label{eigen}
\end{equation}
Then consider $\psi_{N}:=\exp\left(iN\phi\right)\psi$ where $N\in\mathbb{Z}$.
We have that
\begin{equation}
\hat{H}_{\mathrm{tor}}^{K+2N}\psi_{N}=E_{\mathrm{tor}}\psi_{N}
\label{tra}
\end{equation}
as can be seen by using the explicit definition of 
$\hat{H}_{\mathrm{tor}}^{K+2N}$ given in Eq.~(\ref{tor}), together with
the definition of $\psi_{N}$ and Eq.~(\ref{eigen}). 
Eq.~(\ref{tra}) says that $E_{\mathrm{tor}}$ is \emph{also} an energy eigenvalue
of $\hat{H}_{\mathrm{tor}}^{K+2N}$, with $\psi_{N}$ a
corresponding eigenstate. In this way we see that the energy levels
of 
\begin{equation}
\ldots\hat{H}_{\mathrm{tor}}^{-2}, \hat{H}_{\mathrm{tor}}^{0},
\hat{H}_{\mathrm{tor}}^{2}, \hat{H}_{\mathrm{tor}}^{4}\ldots    
\end{equation}
are all the same. Similarly,
\begin{equation}
    \ldots\hat{H}_{\mathrm{tor}}^{-3},
\hat{H}_{\mathrm{tor}}^{-1}, \hat{H}_{\mathrm{tor}}^{1},
\hat{H}_{\mathrm{tor}}^{3}\ldots
\end{equation} 
all have identical energy levels. So, in fact, the energy levels of $\hat{H}_{\mathrm{tor}}^{K}$
are only dependent on whether $K$ is even or odd, and not the particular
value of the integer $K$. In the electromagnetic anology, this can be thought of as a kind of flux quantization condition.

Another property of the Hamiltonian of Eq.~(\ref{tor}) is that, \emph{for $K$ odd, 
all torsional energy levels are doubly degenerate}. 
This is remarkable, as we might expect the torsional potential energy $V$ to cause a splitting of degenerate levels. For $K$ even, this is precisely what happens - levels which are degenerate for $V=0$ are split when $V$ is turned on. 
However, for $K$ odd, the special value of the magnetic flux carried by the solenoid ensures that the degeneracy of torsional energy levels persists even when $V\neq0$. 
This doubling of levels is a known effect in electromagnetic theory \cite{tong2018gauge}
and we give a brief proof here.

Recall that $V$ is symmetric under $\phi\to\phi+\pi$ so $\hat{H}_{\mathrm{tor}}^{K}$ is also symmetric under $\phi\to\phi+\pi$.  As a consequence, we can choose torsional energy eigenstates to be even or odd under $\phi\to\phi+\pi$. Suppose that $E_{\mathrm{tor}}$ is an energy eigenvalue of $\hat{H}_{\mathrm{tor}}^{K}$,
with $\psi$ a corresponding eigenstate which is \emph{even} under $\phi\to\phi+\pi$:
\begin{equation}
\hat{H}_{\mathrm{tor}}^{K}\psi=E_{\mathrm{tor}}\psi.
\end{equation}
Now define a new \wf\ $\tilde{\psi}\left(\phi\right)=\exp(i K\phi)\psi\left(-\phi\right)$.
Then direct calculation, remembering that $V$ is symmetric also under $\phi\to-\phi$, shows that
\begin{equation}
\hat{H}_{\mathrm{tor}}^{K}\tilde{\psi}=E_{\mathrm{tor}}\tilde{\psi}.
\end{equation}
so $\tilde{\psi}$ is \emph{also} an energy eigenstate with the same energy eigenvalue as $\psi$. Note also that 
\begin{equation}
    \tilde{\psi}\left(\phi+\pi\right)=\exp\left(iK\phi+iK\pi\right)\psi\left(-\phi-\pi\right)=\exp\left(iK\pi\right)\tilde{\psi}\left(\phi\right)=\left(-1\right)^{K}\tilde{\psi}\left(\phi\right)
\end{equation}
so, if $K$ is an odd integer, then $\tilde{\psi}$ is \emph{odd} under $\phi\to\phi+\pi$. So we have a pair of degenerate torsional energy eigenstates $\psi$ and $\tilde{\psi}$, which are clearly linearly independent because one is even and the other is odd under $\phi\to\phi+\pi$. In other words, every even eigenstate automatically comes with an odd partner. By the same argument, every odd eigenstate comes with an even partner and so we see that all levels, for $K$ odd, are doubly degenerate.

We will come back to these properties shortly to explain how they correspond 
to well-known properties of the nuclear \wf\ in the presence of conical intersections 
as studied in the molecular Aharonov--Bohm effect. 

In summary, the electromagnetic analogy has given us deeper insight 
into the energy-level structure of the rovibrational states of H$_5^+$. 
In the general case with non-zero $V$, states with $K$ even have energy levels
\begin{equation}
    E=E^{\rm even}_\mathrm{tor}+\frac{1}{2 R^2}J\left(J+1\right) + \left(\frac{1}{2}-\frac{1}{2 R^2}\right) K^2,
    \label{evlev}
\end{equation}
where $E^{\rm even}_{\rm tor}$ takes one of a discrete set of non-degenerate torsional energy eigenvalues. States with $K$ odd have energy levels
\begin{equation}
    E=E^{\rm odd}_\mathrm{tor}+\frac{1}{2 R^2}J\left(J+1\right) + \left(\frac{1}{2}-\frac{1}{2 R^2}\right) K^2,
    \label{oddlev}
\end{equation}
where $E^{\rm odd}_{\rm tor}$ takes one of a discrete set of 
\emph{doubly degenerate} torsional energy eigenvalues. 

H$_5^+$ is an example of a system where the rotational and vibrational dofs
are strongly coupled, so they cannot be treated independently. 
This is the reason why the torsional energy levels depend on the rotational quantum number $K$.
However, we have found that the torsional energy levels only depend on $K$ through whether it is even or odd, and not on its precise value. In particular, they only have to be computed for two values of $K$, say for $K=0$ and $K=1$, and then all the rovibrational energy levels can be deduced. This is obvious from the perspective of the \emph{rovibrational Aharonov--Bohm effect}, as a kind of magnetic flux quantization condition. 
Even more strikingly, every torsional energy level is doubly degenerate for $K$ odd.
Again, this double degeneracy is obvious from our new perspective and 
corresponds to a known electromagnetic effect,\cite{tong2018gauge} 
but it is obscured by the usual decomposition of the rotation-vibration Hamiltonian
into rotational, vibrational, and rovibrational coupling terms.

Before finding the rovibrational energy levels, let us elaborate on the connection 
between the molecular Aharonov--Bohm effect and the rovibrational Aharonov--Bohm effect
we are considering here.

\subsection{Analogy with conical intersections}
For chemists, the most familiar manifestation of the Aharonov--Bohm effect is 
probably the \emph{molecular Aharonov--Bohm effect} (MAB),
in which a nuclear \wf\  encircling a conical intersection of 
potential energy surfaces picks up a minus sign.
In Born--Oppenheimer theory,\cite{27BoOp,54BoHu} one starts by choosing an 
electronic ground-state \wf\  for each nuclear configuration. 
There is a choice of complex phase involved at each nuclear configuration,
since the electronic ground-state is only defined up to a phase. 
Often, we demand that the electronic \wf\  is real-valued (for simplicity,
let us imagine our system is time-reversal invariant and 
that we are ignoring spin-orbit coupling). 
Then, there is still a choice of phase, but there are only two choices and 
they differ by a factor of $-1$ (a phase of $\pi$).

Suppose we choose a real-valued electronic ground-state \wf\  at some nuclear configuration,
and we follow this choice continuously along a closed loop that encircles a conical intersection,
all the time requiring our electronic ground-state \wf\ to be real valued. 
As is well-known, the \wf\ picks up a phase of $\pi$ upon returning to the starting point. 
Single-valuedness of the molecular \wf\ then requires that the nuclear \wf\ also 
pick up a minus sign. 
Another way of characterizing this phase choice for the electronic \wf s is 
that this is the phase choice for which the MTB vector 
potential \cite{mead1979determination} vanishes.

In our nuclear-motion problem, we also have a choice:
for each value of the torsional coordinate $\phi$, 
we have to choose an embedding of the molecule-fixed axes. 
We have been working with the GE throughout this section,
but we are free to transform to some other embedding by specifying a function 
$\theta\left(\phi\right)$ which gives the rotation about the body-fixed $z$-axis
(the symmetry axis of the symmetric top) which relates the old and the new embedding.
Specifying this rotation $\theta\left(\phi\right)$ is much like specifying 
the phase of the electronic ground-state \wf\ in Born--Oppenheimer theory.

So let us start at $\phi=0$ and set $\theta\left(0\right)=0$. 
Now we imagine increasing $\phi$ and, for each $\phi$, 
choosing $\theta\left(\phi\right)$ so that the rotation-vibration coupling vanishes.
Note that this choice is analogous to the choice of a real-valued electronic \wf\ in 
Born--Oppenheimer theory: in the Born--Oppenheimer case,
the MTB vector potential vanishes while in this case,
the vector potential associated with rovibrational coupling vanishes. 
In this way our original GE is transformed to a new embedding, the BE, 
which is depicted in Fig.~\ref{fig:Emb}.

It turns out that when we get to $\phi=2\pi$, the BE embedding has rotated by an angle of
$\pi$ (with respect to the body-fixed $z$-axis) compared to the BE embedding at $\phi=0$.
This rotation by $\pi$ is just like the phase of $\pi$ picked up by a real-valued 
electronic \wf\ upon encircling a conical intersection. 
Now, what does this rotation by $\pi$ mean for quantum states? 
If our rotational state $|JKM\rangle$ has body-fixed angular momentum projection $K$ even,
then a rotation by an angle $\pi$ does nothing to the state and so the state 
is insensitive to this rotation. 
If our rotational state has $K$ odd, however, then it picks up a minus sign. 
In turn, for the rovibrational \wf\ to be single-valued, 
we must have that for $K$ odd the corresponding torsional states pick up a minus sign.
This is just as we found in the previous section: 
elimination of rovibrational coupling is possible, but only if one is prepared to modify
the boundary conditions on the torsional \wf.
With respect to the BE, $K$ even/odd torsional wavefunctions satisfy periodic/anti-periodic boundary conditions. We note that, as a consequence, nuclear motion codes will only yield correct rovibrational energy levels when working in the BE if the boundary conditions are modified compared to the GE.

Thus, the connection between our rovibrational problem and the treatment of 
conical intersections is clearest when we work with the BE. 
Let us temporarily adopt this embedding.
Then we can make two direct connections between our rovibrational problem 
and the familiar treatment of conical intersections. 

First, for $K$ odd, the torsional \wf\  picks up a minus sign 
(like when there is a conical intersection) while, for $K$ even, 
the torsional \wf\  picks up no minus sign (like when there is no conical intersection)
under $\phi\to\phi+2\pi$. 
This is another way of explaining why the torsional energy levels only depend on 
whether $K$ is even or odd, and not on the precise value of $K$. 
It shows that K even/odd just corresponds to periodic/antiperiodic boundary conditions,
\emph{i.e.} conical intersection/no conical intersection. 

Second, it is known that one has to introduce `double groups' to classify nuclear states
in the presence of conical intersections, \cite{althorpe2006general} and 
these larger symmetry groups can lead to higher degeneracies than occur in
the absence of conical intersections. 
This can be viewed as the reason for the double degeneracy of $K$ odd torsional energy levels.
Indeed, the symmetries of the torsional potential $V\left(\phi\right)$ 
under $\phi\to\phi+\pi$ and $\phi\to-\phi$ generate a symmetry group for 
the torsional dynamics isomorphic to $C_{2v}$, 
and this group only has one-dimensional irreducible representations (irreps)
since it is Abelian.
The double group is isomorphic to $C_{4v}$, a non-Abelian group, 
with two-dimensional irreps which correspond to the doubly degenerate levels.

\subsection{Rovibrational energy-level structure \emph{via} first-order perturbation theory}

Now we return to the calculation of the rovibrational energy levels. 
As has been argued, we only need to compute the $K$ odd and the $K$ even torsional levels,
$E^{\rm odd}_{\rm tor}$ and $E^{\rm even}_{\rm tor}$, respectively, in order to deduce 
the full rovibrational energy-level structure [see Eqs.~(\ref{evlev}) and (\ref{oddlev})].
As the torsional potential in the case of H$_{5}^{+}$ is known to be fairly
weak, we can treat the torsional potential by perturbation theory. 

\subsubsection{The $K$ even case}
The $K$ even levels are simply the energy eigenvalues of the $K=0$ torsional Hamiltonian
\begin{equation}
    \hat{H}^{\rm even}_\mathrm{tor} = 2 \hat{p}_{\phi}^2 + \hat{V}.
\end{equation}
We will treat $\hat{V}$ as a small perturbation. 
Ignoring $\hat{V}$, the zeroth order eigenstates can be classified by a $\hat{p}_\phi$ 
eigenvalue $k\in\mathbb{Z}$ and have energy $E^{\rm even}_{\rm tor}=2 k^2$ 
so we get the energy eigenvalues
\begin{equation}
    E^{\rm even}_{\mathrm{tor}}/2=0,1,1,4,4,9,9,\ldots
\end{equation}
For small torsional potentials of the form
\begin{equation}
V\left(\phi\right)=\sum_{j=1}^{\infty}2\tilde{V}_{j}\cos\left(2j\phi\right)=\sum_{j=1}^{\infty}\tilde{V}_{j}\exp\left(2ij\phi\right)+\sum_{j=1}^{\infty}\tilde{V}_{j}\exp\left(-2ij\phi\right),
\end{equation}
where the $\tilde{V}_{j}$ are the Fourier components of the potential,
we have that the matrix element of $\hat{V}$ between states with 
$\hat{p}_\phi$ eigenvalues $k$ and $k'$ is
\begin{equation}
\langle k|V\left(\phi\right)|k'\rangle=\sum_{j=1}^{\infty}\tilde{V}_{j}\left(\delta_{k'+2j,k}+\delta_{k'-2j,k}\right).
\label{matel}
\end{equation}
Thus, in the first order of perturbation theory, the perturbed energies
are 
\begin{equation}
E^{\rm even}_{\mathrm{tor}}/2=0,1-\frac{1}{2}|\tilde{V}_{1}|,1+\frac{1}{2}|\tilde{V}_{1}|,4-\frac{1}{2}|\tilde{V}_{2}|,4+\frac{1}{2}|\tilde{V}_{2}|,9-\frac{1}{2}|\tilde{V}_{3}|,9+\frac{1}{2}|\tilde{V}_{3}|,\ldots
\end{equation}
These are our $K$ even torsional energy levels, where $\tilde{V}_{j}$ diminish quickly
as $j$ increases.
Note that the $E^{\rm even}_{\mathrm{tor}}$ levels are non-degenerate, the potential
has split any degeneracies in the zeroth-order levels.

\subsubsection{The $K$ odd case}
The $K$ odd levels are the energy eigenvalues of the $K=1$ torsional Hamiltonian
\begin{equation}
    \hat{H}^{\rm odd}_\mathrm{tor} = 2 \left(\hat{p}_{\phi}-\frac{1}{2}\right)^2 + \hat{V}.
\end{equation}
Again, we will treat $\hat{V}$ as a small perturbation. Ignoring $\hat{V}$, 
the zeroth order eigenstates can again be classified by a $\hat{p}_\phi$ 
eigenvalue $k\in\mathbb{Z}$ and have energy 
$E^{\rm odd}_{\rm tor}=2 \left(k-\frac{1}{2}\right)^2$ so we get the energy eigenvalues
\begin{equation}
    E^{\rm odd}_{\mathrm{tor}}/2=\left(1/2\right)^{2},\left(1/2\right)^{2},\left(3/2\right)^{2},\left(3/2\right)^{2},\left(5/2\right)^{2},\left(5/2\right)^{2},\ldots
\end{equation}
with $k=0,1,-1,2,-2,3,\ldots$.
Now we introduce $V$ as in the $K$ even case. 
However, there is a crucial difference compared to the $K$ even case.
It is clear from Eq.~(\ref{matel}) that the matrix elements of $V$ only connect states $k$ and $k'$
for which $k-k'$ is an even integer. On the other hand, degenerate zeroth order levels have $k$ and $k'$ related by an odd integer. As a result, the first-order corrections are zero and we get, to first order in perturbation theory,
\begin{equation}
E^{\rm odd}_{\mathrm{tor}}/2=\left(1/2\right)^{2},\left(1/2\right)^{2},\left(3/2\right)^{2},\left(3/2\right)^{2},\left(5/2\right)^{2},\left(5/2\right)^{2},\ldots
\end{equation}

These are our $K$ odd torsional energy levels. 
They are \emph{all doubly degenerate}, the potential does not cause any splitting 
of degeneracies in the zeroth-order levels. 
In fact, we know from our earlier discussion that this degeneracy of 
$K$ odd torsional energy levels is an \emph{exact result}. 
Even if we went to higher order in perturbation theory, 
there would never be any splitting of these levels due to $\hat{V}$.

\begin{figure}
\noindent 
\begin{centering}
\includegraphics[scale=0.425]{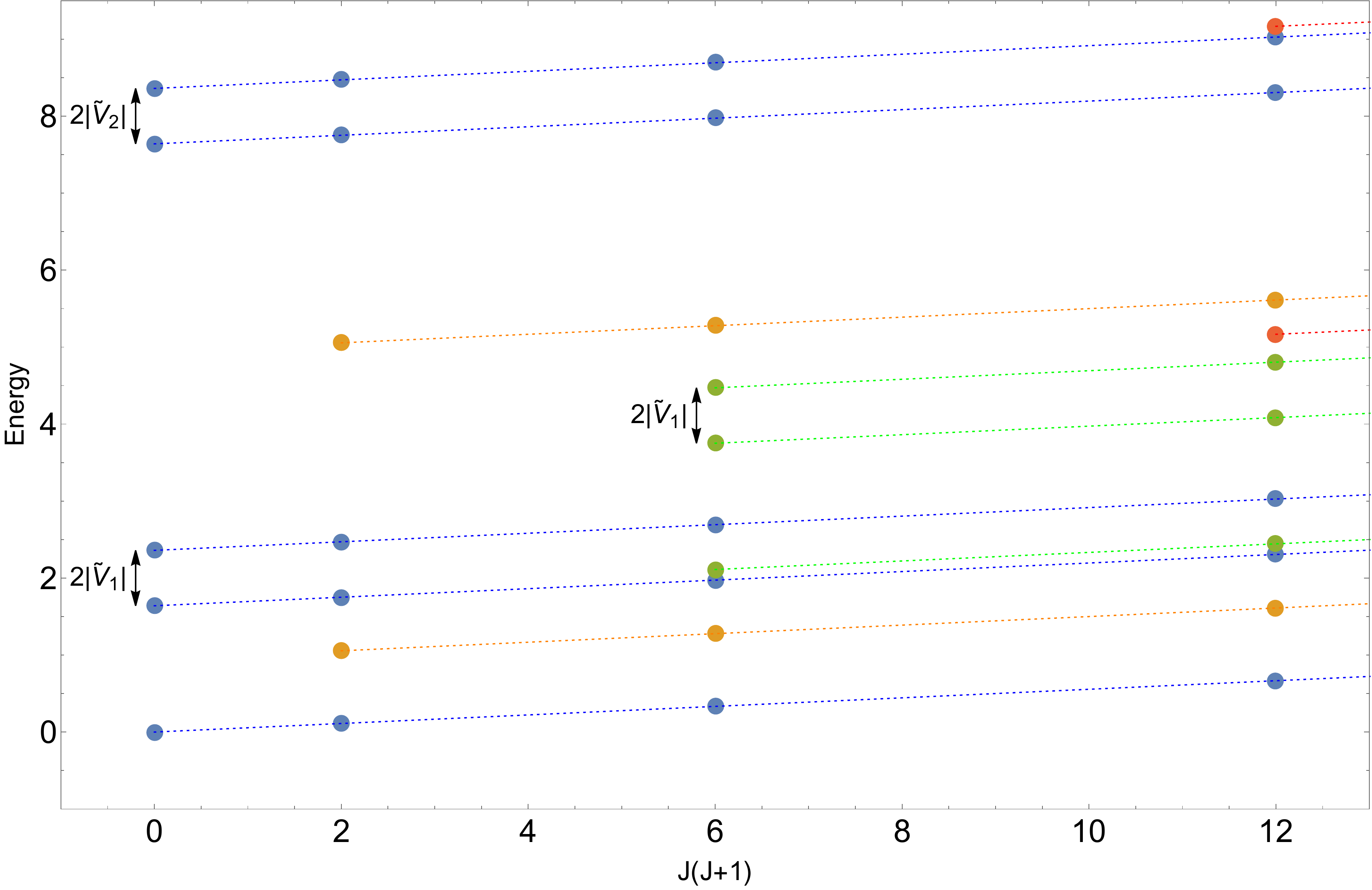}
\par\end{centering}
\caption{Symmetric-top states incorporating torsional splitting. 
The $K=0$ states (blue points) are non-degenerate, the $|K|=1$ states (orange
points) are quadruply degenerate, the $|K|=2$ states (green points)
are doubly degenerate and the $|K|=3$ states (red points) are quadruply degenerate. 
The dotted lines all have gradient $1/2R^{2}$ to illustrate
that the effect of increasing $J$ on the energy levels is simply
to translate the energies (as well as allowing more possibilities
for $K=-J,\ldots,+J$). Level splittings due to the torsional
potential are indicated with arrows.}
\label{fig:sTop}
\end{figure}

The resulting rovibrational energy-level pattern, 
incorporating the rotational kinetic energy $E_{\rm rot}$ 
as well as these torsional energy levels, is depicted in Fig.~\ref{fig:sTop}.
For $K$ odd, the torsional levels are doubly degenerate as we have already explained.
In addition to this torsional degeneracy, there is a further degeneracy under $K\to-K$ 
so that rovibrational levels with $K$ odd are in fact quadruply degenerate 
(this $K\to-K$ degeneracy occurs because the rotational kinetic energy 
only depends on $|K|$, not $K$, while the torsional energy levels are unchanged
under $K\to-K$ since $K$ and $-K$ differ by the even integer $2K$). 
Similarly, rovibrational levels with $K$ even are doubly degenerate 
because of the $K\to-K$ degeneracy. 
The exception to this is when $K=0$, yielding non-degenerate levels.
So, altogether, states with $|K|=0,1,2,3,4,\ldots$ have degeneracies $1,4,2,4,2,\ldots$

\subsection{Beyond the symmetric-top approximation}
Recall that we started our analysis of H$_5^+$ with the 1D($\phi$) torsion model, before making an approximation (the large $R$ limit) which yielded a symmetric-top Hamiltonian. This was a very useful approximation, because the symmetric-top Hamiltonian exhibits the rovibrational Aharonov--Bohm effect and can be understood by analogy with motion in an electromagnetic field. It is also a physically reasonable approximation for H$_5^+$. We now consider what would happen if we dropped this approximation. The answer is that the quadruple degeneracies we have just found would split into $4=2+2$ and the double degeneracies would split into $2=1+1$.

One way of thinking about these splittings is in terms of the symmetries of the symmetric-top rovibrational Hamiltonian in Eq.~(\ref{limit}). Firstly, we have rotations about the symmetry axis of the symmetric top which act on the wavefunction as follows:
\begin{equation}
\hat{R}_{\theta}:\psi\left(\phi\right)\mapsto\exp\left(-i\theta\hat{J}_{z}\right)\psi\left(\phi\right)
\end{equation}
for $\theta\in\left[0,2\pi\right]$. Secondly, we have feasible permutations of the protons, generated by the double transposition
$S=\left(24\right)\left(31\right)$ and the transposition $T_{\pi}=\left(21\right)$, which act on the \wf\ as follows: 
\begin{equation}
\hat{T}_{\pi}:\psi\left(\phi\right)\mapsto\exp\left(-i\pi\hat{J}_{z}\right)\psi\left(\phi-\pi\right)
\end{equation}
\begin{equation}
\hat{S}:\psi\left(\phi\right)\mapsto\exp\left(-i\pi\left(\cos\phi/2\right)\hat{J}_{x}-i\pi\left(\sin\phi/2\right)\hat{J}_{y}\right)\psi\left(\phi\right).
\end{equation}
Finally, we have spatial inversion which is carried out by the parity operator:
\begin{equation}
\hat{P}:\psi\left(\phi\right)\mapsto\exp\left(-i\pi\hat{J}_{x}\right)\psi\left(-\phi\right).
\end{equation}
One can check that $\hat{R}_{\theta}$ and
$\hat{S}$ generate a group $\left\langle \hat{R}_{\theta},\hat{S}\right\rangle \simeq O\left(2\right)$,
while $\hat{R}_{-\frac{\pi}{2}}\hat{T}_{\pi}$ and $\hat{P}\hat{S}$
generate a group $\langle\hat{R}_{-\frac{\pi}{2}}\hat{T}_{\pi},\hat{P}\hat{S}\rangle\simeq D_{8}$ ($D_8$ is the dihedral group of order $8$).
These two groups commute with each other, and together generate the symmetry group
\begin{equation}
G_{\mathrm{symm-top}}\simeq\frac{D_{8}\times O\left(2\right)}{\mathbb{Z}_{2}}.
\end{equation}
Here, the $\mathbb{Z}_{2}$ is generated by a common central element $\hat{R}_{\pi}=\left(\hat{R}_{-\frac{\pi}{2}}\hat{T}_{\pi}\right)^{2}$
of order $2$.

The irreps of $G_{\mathrm{symm-top}}$ have dimensions $1, 2$, and $4$. 
In fact, this is another way of understanding why the rovibrational levels we have found
have degeneracies $1, 2$, and $4$ - they must correspond to irreps of $G_{\mathrm{symm-top}}$. 

If we drop the symmetric-top approximation, then the only symmetries we expect to have left over are those corresponding to spatial inversion, together with feasible permutations of 
the identical nuclei. So we only have the subgroup of $G_{\mathrm{symm-top}}$ which is generated by $\hat{S}$, $\hat{T}_\pi$ and $\hat{P}$. This subgroup is
\begin{equation}
    G\simeq D_8 \times \mathbb{Z}_2
\end{equation}
where the $\mathbb{Z}_{2}$ factor is generated by parity while the
$D_{8}$ factor is generated by the feasible permutations. Note that this group can be identified with $ G_{16}$ in Bunker's notation.\cite{bunker2006molecular}

Clearly $G$ sits in $G_{\rm symm-top}$ as a subgroup:
\begin{equation}
    G\leq G_{\rm symm-top}.
\end{equation}
The idea is that while, to a good approximation, H$_5^+$ is a symmetric top
and has symmetry group $G_{\rm symm-top}$, in reality this symmetry is actually broken down
to the subgroup $G$. The irreps of $G$ have dimensions $1$ and $2$, there are no four-dimensional irreps. Under restriction to the subgroup $G$, 
it turns out that the four-dimensional irreps of $G_{\rm symm-top}$ split into 
a sum $4=2+2$ of two-dimensional irreps of $G$, and that the two-dimensional irreps 
of  $G_{\rm symm-top}$ split into a sum $2=1+1$ of one-dimensional irreps of $G$.
These are the (very small) level splittings we would see amongst the degenerate states 
depicted in Fig.~\ref{fig:sTop} if we took into account the fact that H$_5^+$
is not strictly a symmetric top. 

To illustrate this point further, in Table~\ref{tbl:h5p_roviblevel} 
we reproduce numerical results from Ref.~\citenum{20CsFaSa} for the 
full 1D$\left(\phi\right)$ torsion model, compared with our perturbation theory analysis
of the symmetric-top (Aharonov--Bohm) approximation. 
For this comparison, we need parameter values. 
In the units we have adopted so far in this section, for which $m_H=r=\hbar=1$, we take $\frac{1}{2R^2}=0.0666$, while the torsional potential Fourier coefficients are given
by $\tilde{V}_{1} =0.376$ and $\tilde{V}_{2} = -0.016$,
which corresponds to a torsional barrier of $4\tilde{V}_{1}\approx1.5$.
For the comparisons in Table~\ref{tbl:h5p_roviblevel}, where energies are given in $\textrm{cm}^{-1}$, our energy unit corresponds to $53.34 ~ \textrm{cm}^{-1}$.

\begin{table}
\caption{Torsion-rotation energy levels ($J\geq0$) for H$_5^+$, calculated from the full 1D$\left(\phi\right)$ torsion model and compared to the symmetric-top 
(rovibrational Aharonov--Bohm, RAB) approximation.
The energy levels are given in units of $\textrm{cm}^{-1}$ and all levels up to 
$500 ~ \textrm{cm}^{-1}$ are shown.}
\label{tbl:h5p_roviblevel}
\scalebox{0.9}{
\begin{tabular}{ccr|ccr|ccr|ccr}
\hline \hline
\multicolumn{3}{c}{$J=0$} & \multicolumn{3}{c}{$J=1$} & \multicolumn{3}{c}{$J=2$} & \multicolumn{3}{c}{$J=3$} \\

 1D$\left(\phi\right)$&$\rm{RAB}$&$|K|$&  1D$\left(\phi\right)$&$\rm{RAB}$&$|K|$& 1D$\left(\phi\right)$&$\rm{RAB}$&$|K|$& 1D$\left(\phi\right)$&$\rm{RAB}$&$|K|$ \\
 \hline
0.00 & 0.00  & 0 & 6.68 & 7.10  & 0 & 20.04 & 21.31  & 0 & 40.09 &  42.62 & 0  \\
87.93 & 86.62  & 0 & 55.97 & 56.89  & 1 & 69.16 & 71.10  & 1 & 88.95 & 92.41 & 1  \\
128.26 & 126.74  & 0 & 55.97 &  56.89 & 1 & 69.16 & 71.10  & 1 & 88.95 & 92.41  & 1  \\
429.10 & 425.87  & 0 & 56.14 & 56.89  & 1 & 69.67 & 71.10  & 1 & 89.97 & 92.41  & 1  \\
429.35 & 427.57  & 0 & 56.14 & 56.89  & 1 & 69.67 & 71.10  & 1 & 89.97 &  92.41 & 1  \\
&   &  & 94.61 & 93.73 & 0 & 107.96 & 113.78  & 2 & 127.99 &  135.10 & 2  \\
&   &  & 134.95 & 133.84  & 0 & 113.35 &  113.78 & 2 & 133.37 & 135.10  & 2  \\
&   &  & 273.41 & 270.25  & 1 & 113.36 &  107.94 & 0 & 133.40 & 129.25  & 0  \\
&   &  & 273.41 & 270.25  & 1 & 148.33 & 148.05  & 0 & 168.43 &  169.36 & 0  \\
&   &  & 273.45 & 270.25  & 1 & 201.28 & 200.41  & 2 & 221.31 & 221.72  & 2  \\
&   &  & 273.45 & 270.25  & 1 & 201.28 & 200.41 & 2 & 221.31 & 221.72  & 2  \\
&   &  & 435.78 & 432.97  & 0 & 241.62 & 240.52 & 2 & 261.69 & 221.72  & 2  \\
&   &  & 436.04 & 434.68  & 0 & 241.62 & 240.52 & 2 & 261.69 & 221.72  & 2  \\
&   &  &  &   &  & 286.74 & 284.46 & 1 & 276.08 & 277.36  & 3  \\
&   &  &  &   &  & 286.74 & 284.46 & 1 & 276.08 & 277.36  & 3  \\
&   &  &  &   &  & 286.86 & 284.46 & 1 & 276.08 & 277.36   & 3  \\
&   &  &  &   &  & 286.86 & 284.46 & 1 & 276.08 & 277.36   & 3  \\
&   &  &  &   &  & 449.15 & 447.18 & 0 & 306.74 &  305.77 & 1  \\
&   &  &  &   &  & 449.40 & 448.89 & 0 & 306.74 & 305.77  & 1  \\
&   &  &  &   &  &  &  &  & 306.98 & 305.77  & 1  \\
&   &  &  &   &  &  &  &  & 306.98 & 305.77  & 1  \\
&   &  &  &   &  &  &  &  & 469.20 & 468.49  & 0  \\
&   &  &  &   &  &  &  &  & 469.46 & 470.20  & 0  \\
&   &  &  &   &  &  &  &  & 493.46 & 490.72  & 3  \\
&   &  &  &   &  &  &  &  & 493.46 & 490.72  & 3  \\
&   &  &  &   &  &  &  &  & 493.47 & 490.72  & 3  \\
&   &  &  &   &  &  &  &  & 493.47 & 490.72  & 3  \\

\hline \hline
\end{tabular}
}
\end{table}

Note that $K$ is no longer a good quantum number in the 1D$\left(\phi\right)$ model, 
but because the symmetric-top symmetry is only slightly broken,
it is still useful to label states by $|K|$, as seen in Table~\ref{tbl:h5p_roviblevel}.
The small splittings $4=2+2$ and $2=1+1$ due to the breaking of the 
symmetric-top symmetry are evident in Table~\ref{tbl:h5p_roviblevel}. 
For example, looking at the $J=3$ column, we see that the first row corresponds to 
a non-degenerate $K=0$ level at $42.62 ~ \rm{cm}^{-1}$ in the symmetric top approximation.
The next four rows correspond to a quadruply degenerate $|K|=1$ level 
at $92.41 ~ \rm{cm}^{-1}$ in the symmetric top approximation,
which is split as $4=2+2$ into doubly degenerate levels at 
$88.95 ~ \rm{cm}^{-1}$ and $89.97 ~ \rm{cm}^{-1}$. 
The next two rows after this correspond to a doubly degenerate $|K|=2$ level 
at $135.10 ~ \rm{cm}^{-1}$ in the symmetric top approximation, 
which is split as $2=1+1$ into non-degenerate levels at 
$127.99 ~ \rm{cm}^{-1}$ and $133.37 ~ \rm{cm}^{-1}$.

\section{Conclusions}
By revisiting the problem of coupling overall rotations with an internal vibrational motion
for symmetric-top molecules,
we demonstrated that this coupling can be understood by analogy 
with the famous Aharonov--Bohm effect, with the rotational motion influencing 
the vibrational motion in a way anologous to the influence of a magnetic solenoid 
on a charged particle. The effective electric charge is identified with $K$, 
the component of the angular momentum along the symmetry axis of the molecule, 
while the effective magnetic flux carried by the solenoid is proportional to $\alpha$,
the rovibrational coupling strength. In particular, the quantum energy levels associated with the vibrational motion are affected by the rovibrational coupling in a characteristic way familiar from electromagnetic theory.

As an application, we considered the low-energy rovibrational dynamics of H$_5^+$,
which are known from previous studies to be well-described by a
model coupling overall rotations to a single torsional motion. 
We showed that, since H$_5^+$ is approximately a symmetric top, 
the rovibrational Aharonov--Bohm effect governs the low-energy dynamics 
of the ion to a good approximation. 
Moreover, in the case of H$_5^+$, the value of the effective magnetic flux carried by the solenoid within this analogy 
is such that the torsional energy levels only depend on whether $K$ is even or odd,
and that all torsional energy levels for $K$ odd are doubly degenerate. 
These latter effects have analogues in the well-known molecular Aharonov--Bohm effect,
in which the presence of a conical intersection of Born--Oppenheimer potential 
energy surfaces influences the nuclear dynamics through the Mead--Truhlar--Berry
vector potential.

The electromagnetic analogy gives us a new way of understanding 
the rovibrational level structure of symmetric-top molecules, 
as we have vividly demonstrated for H$_5^+$. 
This gain in understanding is largely due to the fact that changes of embedding of molecule-fixed axes 
play a transparent role in the electromagnetic analogy, corresponding to gauge transformations.
This is to be contrasted with the usual decomposition of the nuclear Hamiltonian 
into vibrational, rotational and rovibrational terms, in which the behavior 
under changes of embedding is somewhat obscured. 

A natural question is whether these ideas can be extended to the case of asymmetric-top molecules.
This can be done, but one has to replace the effective magnetic field 
which couples to the vibrational dynamics with a so-called non-Abelian gauge field.
This perspective on general rotation-vibration coupling is a well-developed 
subject,\cite{Shapere1987,Shapere1989,GUICHARDET1984,Littlejohn1997}
which in a sense reduces to the electromagnetic analogy introduced here 
in the special case of a symmetric top. 
An attractive feature of the electromagnetic analogy is that the mathematical ideas involved
are already familiar to chemists in the context of the molecular Aharonov--Bohm effect.
The non-Abelian fields which arise in the more general case are not as familiar,
although we point out that they do in fact have analogues in the chemistry literature.
For example, in molecular systems with strong spin-orbit coupling 
one may be interested in the intersection of two potential energy surfaces,
each surface corresponding to a doubly (Kramers) degenerate electronic state.
This picture naturally leads to the introduction of a non-Abelian gauge field 
which couples to the nuclear dynamics, as explored in 
Refs. \citenum{mead1980electronic}, \citenum{mead1987molecular} and \citenum{johnsson19972}.
We expect that the electromagnetic analogy, and its non-Abelian generalization,
will continue to yield new insights in nuclear dynamics.

\section*{Acknowledgements}
The work performed 
received support from NKFIH (grants no. K119658 and K138233) 
and from the ELTE Institutional Excellence Program (TKP2020-IKA-05).

\section*{Conflicts of interest}
There are no conflicts to declare.

\FloatBarrier     		
{\small
\bibliography{H5plus}}

\end{document}